\documentclass[conference]{IEEEtran}
\IEEEoverridecommandlockouts
\usepackage{cite}
\usepackage{amsmath,amssymb,amsfonts}
\usepackage{algorithmic}
\usepackage{graphicx}
\usepackage{textcomp}
\usepackage{xcolor}
\usepackage{flushend}
\usepackage{dblfloatfix}
\usepackage{paralist}
\def\BibTeX{{\rm B\kern-.05em{\sc i\kern-.025em b}\kern-.08em
    T\kern-.1667em\lower.7ex\hbox{E}\kern-.125emX}}
\begin{document}

\title{Realizing Forward Defense in the Cyber Domain$^*$\\
\thanks{$^*$Invited Position Paper. Distribution Statement A: This material is based upon work supported by the Under Secretary of Defense for Research and Engineering under Air Force Contract No. FA8702-15-D-0001. Any opinions, findings, conclusions or recommendations expressed in this material are those of the author(s) and do not necessarily reflect the views of the Under Secretary of Defense for Research and Engineering. © 2021 Massachusetts Institute of Technology. Delivered to the U.S. Government with Unlimited Rights, as defined in DFARS Part 252.227-7013 or 7014 (Feb 2014). Notwithstanding any copyright notice, U.S. Government rights in this work are defined by DFARS 252.227-7013 or DFARS 252.227-7014 as detailed above. Use of this work other than as specifically authorized by the U.S. Government may violate any copyrights that exist in this work.}
}

\author{\IEEEauthorblockN{Sandeep Pisharody, Jonathan Bernays, Vijay Gadepally, Michael Jones, \\ Jeremy Kepner, Chad Meiners, Peter Michaleas, Adam Tse,  Doug Stetson}
\IEEEauthorblockA{MIT Lincoln Laboratory, Lexington, MA, USA\\
}}

\maketitle

\begin{abstract}
With the recognition of cyberspace as an operating domain, concerted effort is now being placed on addressing it in the whole-of-domain manner found in land, sea, undersea, air, and space domains. Among the first steps in this effort is applying  the standard supporting concepts of security, defense, and deterrence to the cyber domain. This paper presents an architecture that helps realize forward defense in cyberspace, wherein adversarial actions are repulsed as close to the origin as possible. However, substantial work remains in making the architecture an operational reality including furthering fundamental research cyber science, conducting design trade-off analysis, and developing appropriate public policy frameworks.
\end{abstract}


\section{Introduction}
    Current cyber network defenses are often focused on implementing a perimeter defense (firewall) to block expected malicious traffic, by simultaneously monitoring traffic that crosses the firewall and identifying and mitigating cyber-attacks. This approach puts cyber defenders at a significant disadvantage, since the `fight' between defenders and attackers is almost always on the defenders  cyber terrain (blue space), or very close to it. Furthermore, this firewall approach offers zero to limited advanced knowledge of attacks since observations are from the limited vantage point of one’s own network, which is unlikely to yield sufficient information to isolate attackers and respond with appropriate defensive cyber operations.

    The observe-pursue-counter defend forward approach is the foundation of defense systems in many domains \cite{deeks2004defend}\cite{kosseff2019contours}. The defend forward approach in cyber, as defined in the Cyberspace Solarium Commission report \cite{montgomery2020cyberspace}, calls for the collection and aggregation of traffic to identify malicious activity, making any and all actions dependent on observability. It is only after the observe component is fully understood, that we can lean on the detailed detection requirements driven by pursue-counter elements \cite{gold1969digital}\cite{kay2013fundamentals}\cite{oppenheim1999discrete}. Thus, in many deployments, the observe component is the most technologically challenging and resource intensive. 

    Broad data collection from multiple collection points provides a much more comprehensive view of concurrent IP traffic, and  makes it possible to produce synoptic views with meaningful insights. Consider, for example, botnet activity targeted at blue space systems. Patterns of bot activity may be either unobserved or less obvious from a purely local vantage point, and will not be apparent without a broader view. Furthermore, analysis of broad data collection helps identify attacks that may take place against distant targets, and analysis of those attacks can provide guidance to the Cyber Mission Teams (CMTs) to be proactive in defense of blue space.

    In light of this observation, we have created a reference architecture for a prototype that seeks to conduct analysis of multiple broad-aperture collections of network data to detect adversarial activity -- thereby advancing Observe capabilities to enable the Pursue component of the defend forward approach. The capability, named CyPhER (Cyber Phenomenology Exploitation and Reasoning), uses substantial and targeted pruning of large packet captures along with mathematical and computational advancements upending the broadly accepted notion that a synoptic view of adversarial events is untenable to achieve an over-the-firewall-horizon (OTFH) defense capability.  Long range detection is enabled by focusing traffic characterization and anomaly detection on network traffic represented only as anonymized source/destination IP (Internet Protocol) pairs that are readily obtainable and uniquely provide observations of networks over largest temporal and spatial scales.
    
    The remainder of this document specifies system architecture and the building blocks to a robust, cost-effective tool that can provide a synoptic view of adversarial events in global IP traffic.  Throughout the design  privacy is a primary goal and it should be assumed that all data in the system are anonymized unless other specified. The counters enabled from having this synoptic view can be customized according to mission needs, and can range from forensics, to early warning, and even cyber deterrent actions. The architecture is general in nature, with specific design choices highly dependent on the mission and the mission concept of operations (CONOPS).
\section{Motivation}
    
    The ability to continuously monitor a complete operating domain is a critical element of defense systems in most domains. Its widely acknowledged that the absence of broad aperture data analysis  puts defenders at an asymmetric disadvantage allowing adversaries to amass resources undetected to target the most vulnerable points. The ability for defenders to get situational awareness, and get synoptic views of what is happening goes a long way in being able to defend strategic assets \cite{endsley1995toward}\cite{barford2010cyber}.
    
    Consider the air defense sector for example. In the 1930s, the prevailing view that the ``bomber always get through'' focused air defense on costly preemptive strikes and counterstrikes requiring massive bomber fleets.  By the 1940s, long-range radars and integrated air defense had changed the game \cite{chernyak2009brief}, and were used to detect incoming bombers before they reached the protected air space, buying defenders valuable time to orient and react, and possibly prevent the breach of air space. We posit that cyber defense has a lot in common with air defense in the 1930s, and there is a need for the appropriate ``radar'' to enable an integrated cyber defense system that enables specific actions to threats detected before the threat is within our perimeter. 
    
    While the ability to collect and process massive quantities of cyber data has long been considered a roadblock to wide aperture data cyber data analysis, it is no larger than the problems we routinely solve on our supercomputers today in other domains. Through advances made in matrix mathematics \cite{kepner2016mathematical}, super computing and insights from some fundamental cyber phenomenology \cite{kepner2019new}, we now have the ability to create a tool that can dramatically increase the scope of cyber sensing and enable left-of-launch defense strategies for cyber. The ability to see into grey cyberspace will enable us to know of adversarial actions against targets outside of blue space, and will help inform defenses. 
\section{Cyber Forward Defense Vision}
\label{sec:forward}

    While cyberspace has been treated as an operating domain for years, addressing it in a whole-of-domain manner such as in land, sea, undersea, air, and space is more recent development\cite{kepner2021zero}. Consistent with other domains, standardizing cyberspace operating domain begins with applying the three supporting elements that comprise domain protection: security, defense, and deterrence. As defined in the Dictionary of Military and Associated Terms \cite{gortney2021dictionary} for the cyber domain, these are:
    \begin{itemize}
        \item Cyberspace Security: Actions taken within protected cyberspace to prevent unauthorized access to, exploitation of, or damage to computers, electronic communications systems, and other information technology, including platform information technology, as well as the information contained therein, to ensure its availability, integrity, authentication, confidentiality, and nonrepudiation.
        \item Cyberspace Defense: Actions taken within protected cyberspace to defeat specific threats that have breached or are threatening to breach cyberspace security measures and include actions to detect, characterize, counter, and mitigate threats, including malware or the unauthorized activities of users, and to restore the system to a secure configuration.
        \item Cyberspace Deterrence: The prevention of action by the existence of a credible threat of unacceptable counteraction and/or belief that the cost of action outweighs the perceived benefits.
    \end{itemize}
        
    \begin{figure}[!htb]
        \centerline{\includegraphics[width=0.5\textwidth]{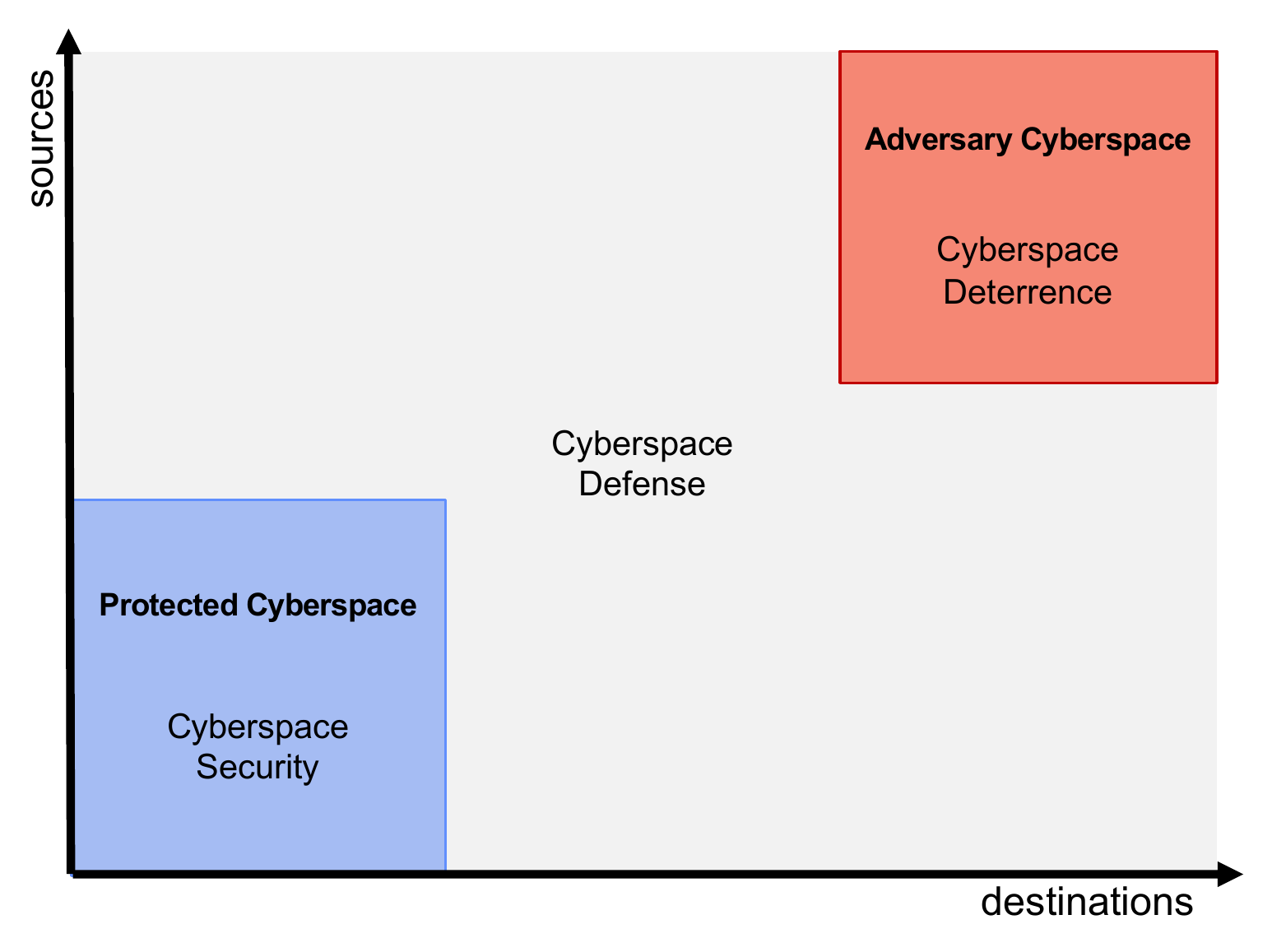}}
        \caption{Source-destination traffic matrix view and associated standard domain cyberspace terminology.}
        \label{fig:matrix}
    \end{figure}
    
    A critical element of domain defense systems analysis is creating an effective picture of the operating domain that is both understandable by decision makers and accurate to implementors.  By condensing network traffic to source-destination IP pairs, we can use an xy-plane to represent all Internet traffic at a moment in time \cite{kepner2021zero}. Assuming the lower values on the x- and y-axis represent internal IPs and the higher values on the x- and y-axis represent adversarial red-space, we can use Figure~\ref{fig:matrix} to visually separate areas where cyberspace security, defense, and deterrence come into play.
    
    \begin{figure}[!htb]
        \centerline{\includegraphics[width=0.5\textwidth]{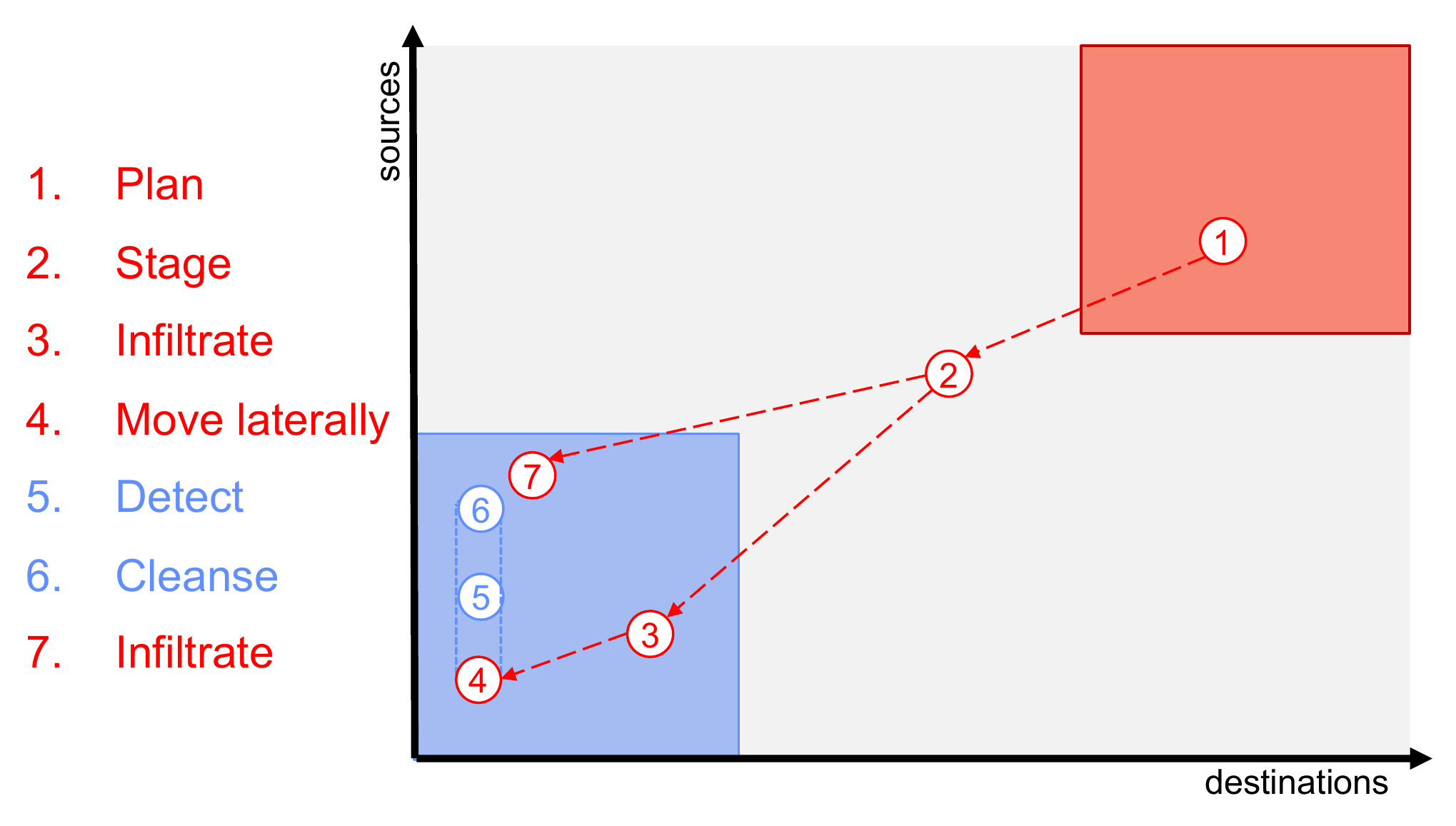}}
        \caption{Notional cyber-attack detected (5) and cleansed (6) well after initial infection (3).}
        \label{fig:attack}
    \end{figure}
    
    Figure~\ref{fig:attack} represents a notional adversarial cyber attack using Lockheed Martin's Cyber Kill Chain \cite{hutchins2011intelligence} mapped to the traffic matrix view from Figure~\ref{fig:matrix}. The threat model being considered in this notional scenario is a widely observed attack pattern. 
    Reconnaissance and weaponization, the first two stages in the kill chain correspond to planning in the adversarial domain (1). The delivery stage corresponds to staging in neutral space (2). The exploitation stage of the kill chain follows infiltration into the blue space as shown in (3). Once inside a protected domain, spreading begins and expands the footprint of adversarial capability (4) allowing for the remaining three kill chain stages: installation, command and control (C2), and actions on objectives. Any blue space action (5, 6) to limit discovered adversary operations often do little to prevent the adversary from continuing to explore further attack paths as shown in (7).
    
\begin{figure}[!ht]
        \centerline{\includegraphics[width=0.5\textwidth]{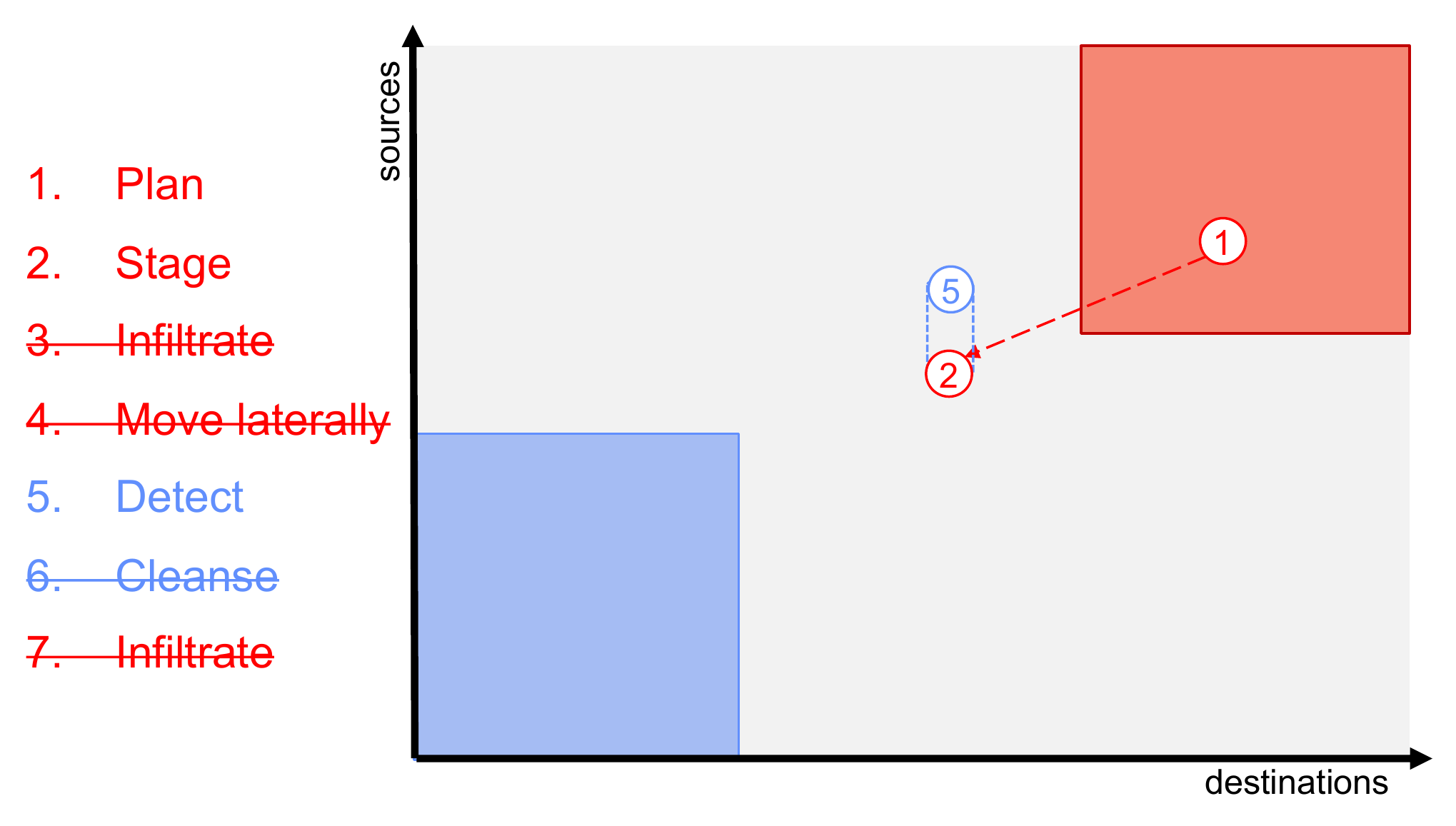}}
        \caption{Notional cyber-attack detection (5) with defend forward techniques.}
        \label{fig:future}
    \end{figure}

    Reducing the time to detect ($t_{\rm detect}$) greatly enhances the effectiveness of any domain defense system. Techniques that move surveillance to IP space outside of protected enclave  not only shortens $t_{\rm detect}$, but also provide cyber defenders more lead time to secure blue cyber terrain prior to infiltration by adversary. Figure~\ref{fig:future} shows the CyPhER vision for such a defensive system. While such an architecture is easy to hypothesize, a detailed systems analysis is required to investigate the practical feasibility of such an approach. Section~\ref{sec:arch} lays out a functional decomposition of what such a system could look like.
    
\section{Architectural Framework}
\label{sec:arch}

    This Section discusses the functional decomposition of the CyPhER cyber defense system that can serve as the reference for future prototypes. The selected framework is based on an open design architecture that emphasizes flexibility to adapt to multiple mission needs. An open design architecture emphasizes publicly shared design information that uses widely available software where the final product is shaped by the mission and end users \cite{bessen2016knowledge}\cite{howard2012open}\cite{pearce2012new}. Furthermore, such an architecture can address the needs of interfacing models from multiple distinct organizations that might be responsible for different portions of the mission.
    
    \begin{figure*}[btp]
        \centering
        \includegraphics[width=\textwidth]{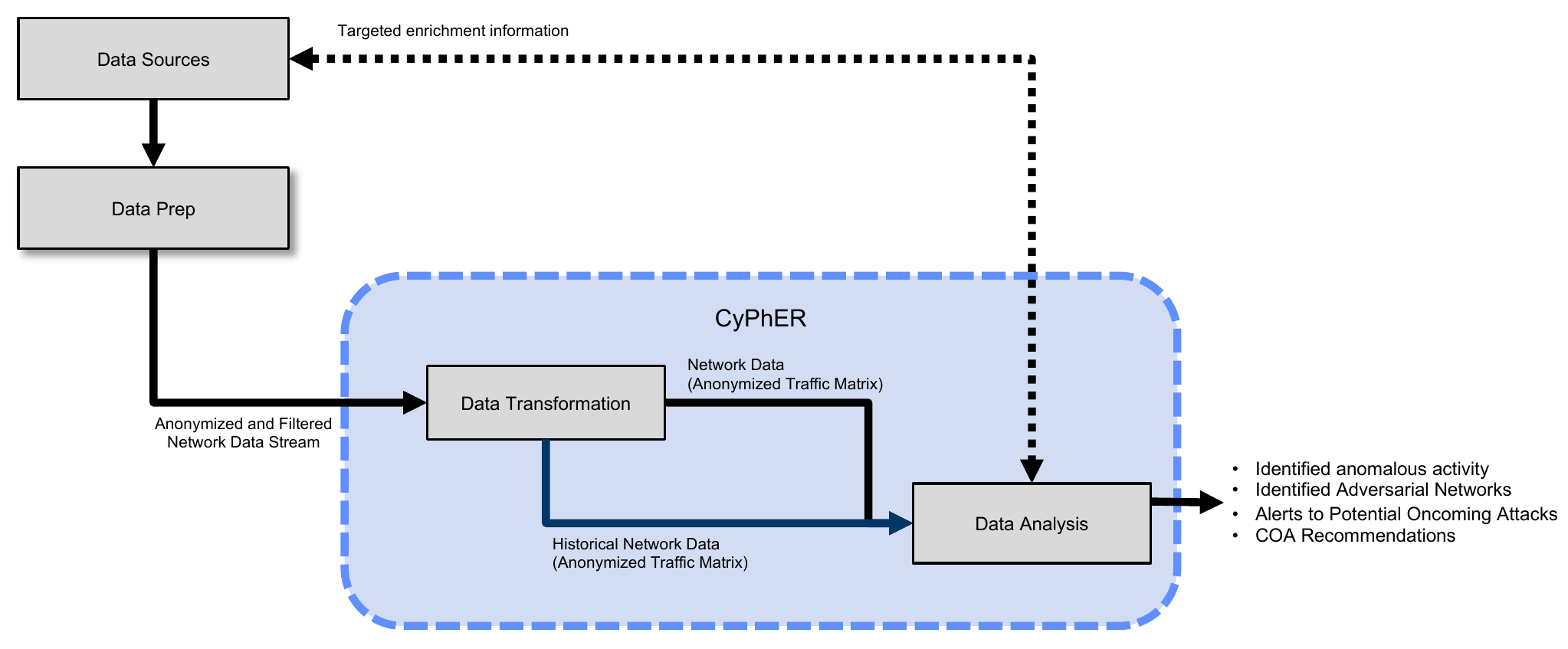}
        \caption{High-level architecture framework for CyPhER cyber defense system.}
        \label{fig:framework}
    \end{figure*}
    
    The architectural framework for the prototype is depicted in Figure~\ref{fig:framework}. Since the architecture aims to allow for a prototype to consume myriad data sources in a fast and inexpensive manner; the architectural components are loosely coupled to provide the most flexibility for updating individual components without completely tearing down the ingestion process. Specific product recommendations (COTS, GOTS, or custom) are dependent on mission needs and are outside the scope of this paper.
    
    \subsection{Data Sources}
    \label{sec:src}
        To enable the prototype to be broadly deployable, the architecture is able to accept multiple internal and external data feeds. Since these data sources might have different rates of traffic and different levels of fidelity, it is important to filter out the relevant pieces of information and curate/homogenize inputs for processing~\cite{sawant2013big}. This seamless merging and consolidation enables the analytics engines to process the disparate sources as a single, large, homogenized dataset. 
    
\begin{figure}[]
            \centerline{\includegraphics[width=0.5\textwidth]{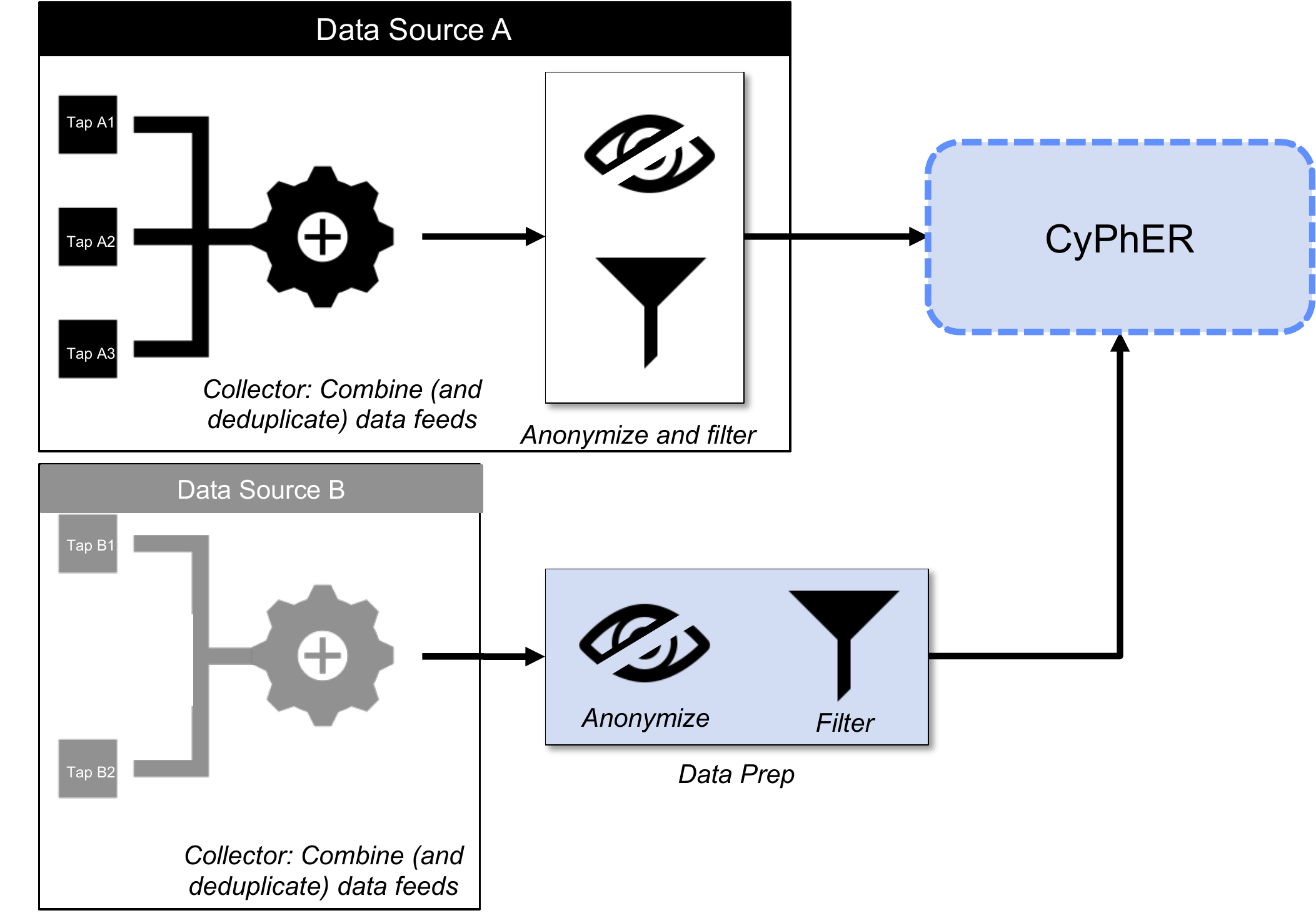}}
            \caption{Collection, anonymization, and filtering of data feeds prior to analysis using the CyPhER prototype.}
            \label{fig:ingest}
        \end{figure}

        The filtering process drops all but four key fields of information about each packet: anonymized source IP, anonymized destination IP, timestamp, and tap location. This dimensionality reduction reduces the amount of data to be processed, and opens up the possibility of analyzing network traffic at the scale of global Internet traffic, as demonstrated in \cite{kepner202075}. 
        Analysis of compute capacities required to analyze global Internet traffic is presented in \cite{kepner2021zero}.
        The filtered data fields, source and destination IPs, are anonymized at the source prior to ingestion into the data transformation module as shown for ``Data Source A" in Figure~\ref{fig:ingest}. In cases where the data sources are raw, unanonymized, unfiltered packet captures, or NetFlow without anonymization, an optional data prep service can be used to accept the data feed and anonymize/curate the raw feed as needed, as shown for ``Data Source B" in Figure~\ref{fig:ingest}. 

        Much of the value proposition of this work is realized when network traffic data from multiple collection points, often controlled by different organizations, are aggregated and analyzed (discussed in Section~\ref{sec:tap}).  While anonymization is not a requirement for the analysis algorithms to be effective, there is tremendous value in privacy-preserving anonymization, since it alleviates many data sharing concerns between organizations.  The use of traffic matrix representations and matrix based algorithms allows analysis to work independent of anonymization, greatly simplifying the overall design and implementation, while also enabling privacy.
        
    \subsection{Data Transformation}
\begin{figure}[]
            \centerline{\includegraphics[width=0.5\textwidth]{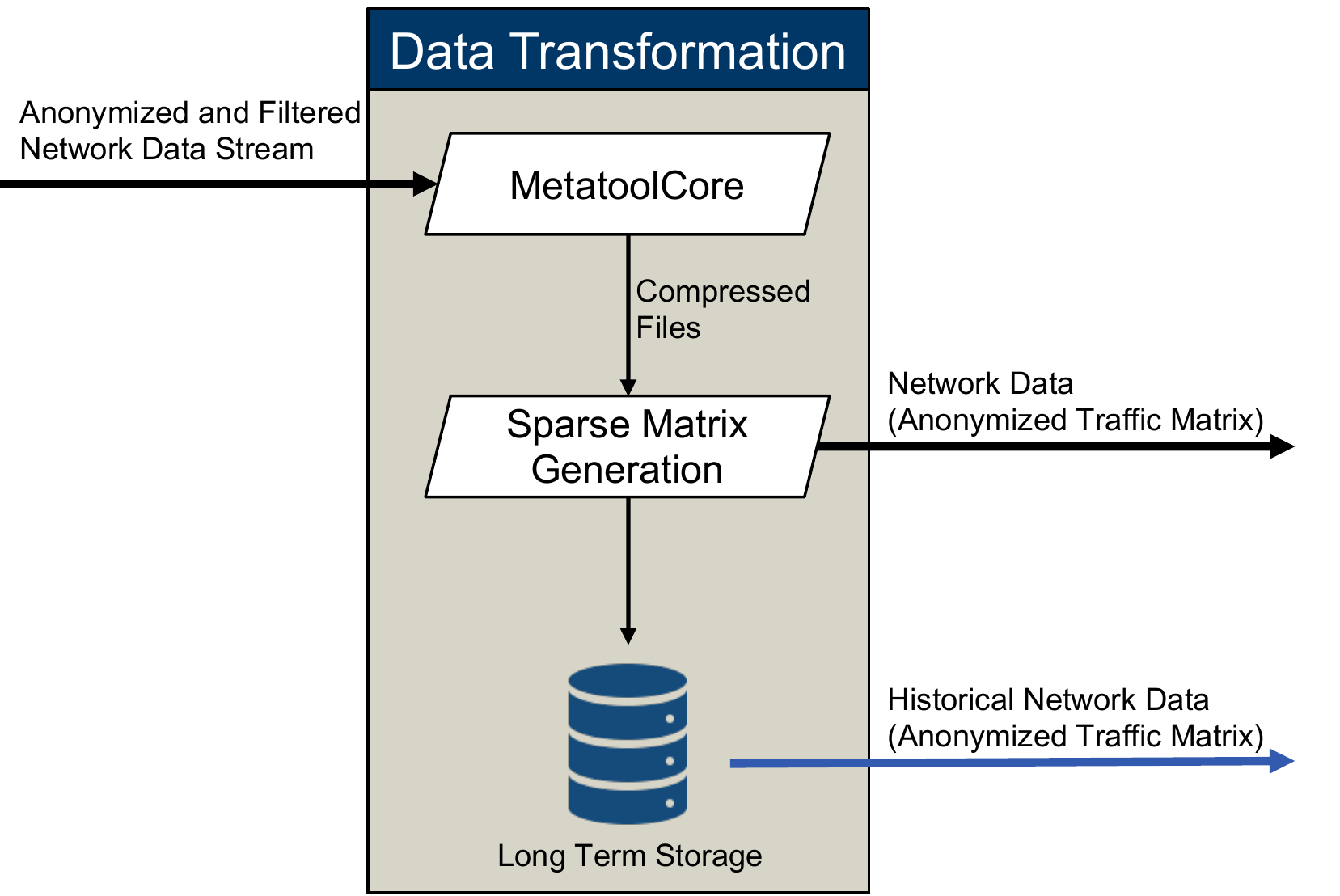}}
            \caption{Data transformation module of CyPhER prototype converts the anonymized and filtered network data stream to anonymized network traffic matrices for long term archival and statistical analysis.}
            \label{fig:transform}
        \end{figure}

        The data transformation module accepts incoming data feeds in a number of different formats (pcap, csv, binary, etc.) and transforms them into compressed files using a custom-built MetatoolCore code suite, the functionality of which can be implemented by any number of widely available capabilities. These compressed binaries can be 
        \begin{inparaenum}[a)]
            \item stored for long term archival on storage systems that are expressive enough to store streams of matrices that represent timestamped graphs with nodes and edges evolving over time; and 
            \item converted to anonymized network traffic matrices for processing using the GraphBLAS \cite{graphblas} network analysis package. 
        \end{inparaenum}
        Figure~\ref{fig:transform} shows the functional components of the data transformation module.
        
    \subsection{Data Analysis}
    \label{sec:analysis}
\begin{figure}[]
            \centerline{\includegraphics[width=0.5\textwidth]{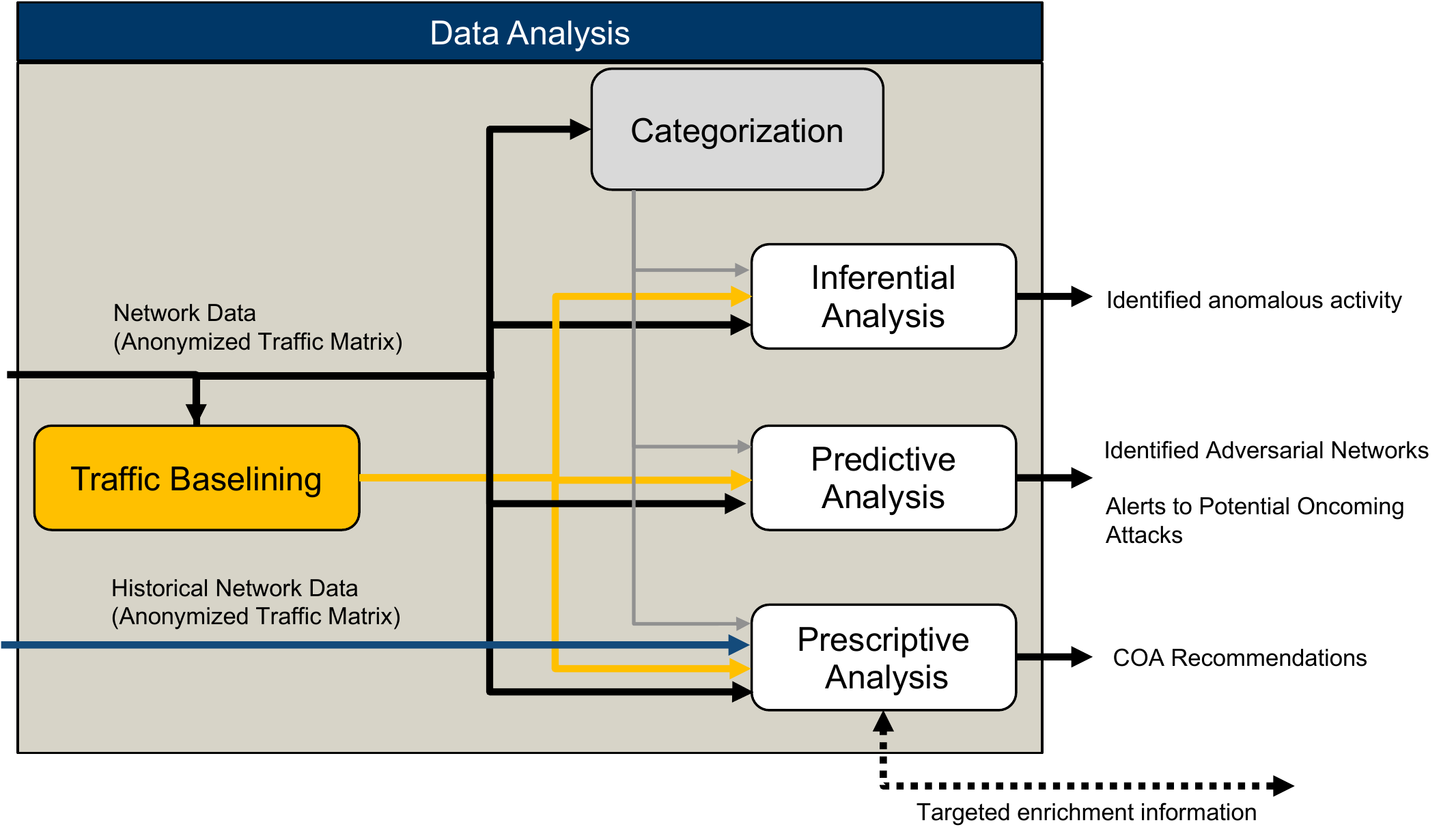}}
            \caption{Data analysis module in the CyPhER architecture.}
            \label{fig:analysis}
        \end{figure}

        The data analysis module is central to the unique value offered by the CyPhER cyber defense capability. This module takes in as input the anonymized and filtered data streams, and computes traffic baselines (including variances from the mean). The traffic baselines are encoded as coefficients of a mathematical model representing ``normal'' traffic, and are continuously updated with new incoming network data. The baseline traffic is used to inform four different types of data analyses, as shown in Figure~\ref{fig:analysis}:
        \begin{itemize}
            \item Categorization, which sorts IP address into clusters based on their graph attributes. This type of analysis detects scanning IPs, darkspace IPs, and nominally active IPs and serves as foundational data for the other types of analyses.
            \item Inferential analysis, or statistical study of network data streams. This sub-module analyzes the incoming network streams to detect deviations from expected traffic patterns as gleaned from traffic baselines. While the architecture maintains flexibility in selecting the specific solution to accomplish this task, the GraphBLAS \cite{kepner202075}\cite{graphblas} package is highly recommended because of its powerful parallel graph algorithm and hypersparse matrix capabilities.
            \item Predictive analysis, wherein along with the network data and traffic baselines, historical data from long term storage is used to identify adversarial infrastructure and predict oncoming attacks. The predictive analysis module leverages AI/ML techniques to identify clusters of IPs that behave similarly, as well as detect patterns of activity that precede known historical adversarial activity to warn of oncoming attacks that follow similar strategy.
            \item Prescriptive analysis, which recommends courses of action (COAs) to counter specific threats in line with the persistent engagement doctrine \cite{command2018achieve}\cite{smeets2020us}. While many of these COAs can be generated using anonymized data, a communication channel with the collection point for selective, targeted enrichment is available if needed. This allows for specific subsets of the data to be deanonymized as needed, while limiting it to determining COAs. Alternatively, the anonymization technique could be selected after demonstrating that permitted COAs can be generated using anonymized data.
        \end{itemize}
        
        While entirely customizable, the outputs from categorization analysis are likely to be utilized by the inferential, predictive, and prescriptive analysis sub-modules. However, data exchange between other sub-modules are dependent on the data separation policies and procedures in place between organizations tasked with inferential, predictive, and prescriptive actions.
\section{Design Tradeoffs and Challenges}
\label{sec:challenges}

    The open design architecture discussed in Section~\ref{sec:arch} provides the framework for implementing an observe-pursue-counter capability using the CyPhER prototype. This Section addresses some of the tradeoffs that need to be considered with respect to addressing critical parameters such as the number of data sources, locations of data taps, sampling intervals, format of data to be collected and processed, etc.

    \subsection{Tap Location}
    \label{sec:tap}
\begin{figure}[htbp]
            \centerline{\includegraphics[width=0.5\textwidth]{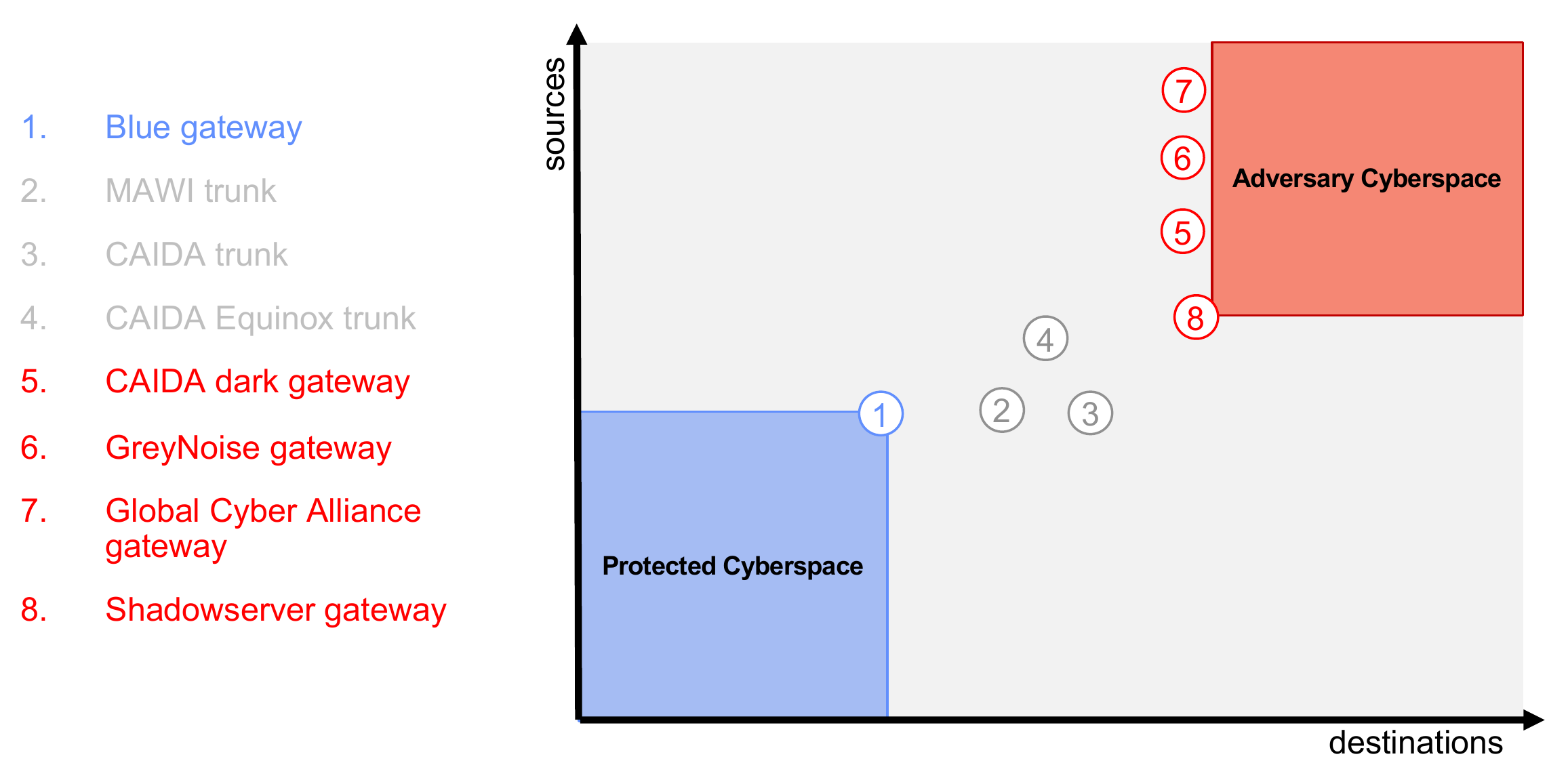}}
            \caption{Examples of current observatories and outposts and their approximate proximity to various network traffic.}
            \label{fig:outposts}
        \end{figure}

        There are a number observatories and outposts in operation today. These sites are a mixture of academic, non-profit, and commercial efforts and provide different viewpoints into the network landscape, as shown in Figure~\ref{fig:outposts}. Data from gateways of protected space are commonly available to most organizations. Others such as data sets from Center for Applied Internet Data Analysis (CAIDA) and Measurement and Analysis on the WIDE Internet (MAWI) lie along major trunk lines in grey space. And some are honeypots (GreyNoise), or dark spaces (unassigned locations on the Internet) that see mostly adversarial traffic (CAIDA Telescope), or sunkholed botnet command-and-control servers (ShadowServer).
        
        Collection points (taps) exist today to sample traffic across major portions of the entire internet, via high bandwidth taps at strategic locations accessible to ISPs. Such taps make it feasible to collect the data required as input into CyPhER. 
        
        One of the primary considerations that needs to be addressed is the placement of taps that enable the necessary sampling of traffic to utilize the adversarial activity detection capability of CyPhER. However, no deterministic method exists that helps decide where taps should be placed. Placement of taps closer to the blue network, will capture more targeted data (either source or destination in the blue network). However, this is more limiting and provides less time to react and less of a synoptic view of adversarial activity in the overall topology. Placing the taps in grey space, or in red-adjacent space can capture a lot more data, much of it not destined for the blue space. However, these taps have the potential to help capture additional anomalous phenomenon by identifying adversarial activity patterns that are subtly different and target a different destination, thereby enabling the CyPhER prototype to be able to thwart similar activity directed toward blue space.

        Combining multiple correlated data sources from different collection points away from blue space has the potential to provide a live synoptic view of traffic, expanding observability, as well as providing forensic benefits through identifying if
        \begin{inparaenum}[a)]
            \item patterns of adversarial activity happened elsewhere; and
            \item adversarial tactics changed.
        \end{inparaenum}
        
        Much like the aviation sector, where integrating sensor outputs from multiple high-resolution local weather sensors provides wide-area, high-resolution weather information that has improved the safety and efficiency of air travel, we believe capturing IP traffic from multiple data collection points can have a similar impact. Much of the benefits of such analysis can be gained by utilizing existing network traffic taps around IP spaces of interest instead of deploying new collection points on backbone routers.

    \subsection{Number of Data Sources}
        An additional challenge is determining the number of disparate data sources required as an input to the CyPhER system for effective detection of adversarial threats. Initial observations suggest that a modest number of collection points can provide a synoptic view of global IP traffic.
        
        While the architectural framework presented in Section~\ref{sec:arch} is scalable, and can process multiple input data streams, the number of data streams required is dependent on
        \begin{inparaenum}[a)]
            \item the accuracy and precision desired from the detector; and 
            \item the correlation and/or overlap in the data between the collection sources.
        \end{inparaenum}
    
    \subsection{Data Format and Privacy}
        The extensibility and open design architecture being adopted for the CyPhER prototype enables us to change the types of data used for system inputs. While the algorithms currently used to detect anomalies use deviations from baselines derived from matrices of  source-destination IP pairs, any added benefits provided by using additional header fields for anomaly detection needs to be examined. Any additional value provided in adversarial activity detection or accuracy and precision metrics would have to be weighed against the additional cost of storing and processing multiple additional fields of data.
        
        Additionally, as mentioned in Section~\ref{sec:src}, there is tremendous value in privacy-preserving anonymization, since it alleviates many data sharing concerns between organizations. Anonymization, however, is not a requirement for the analysis algorithms to be effective. Analysis of how much information is obscured by the privacy-protection processes, and what limitations are imposed by working solely with anonymized data needs to be examined in detail.
\section{Next Steps}

    There are several concrete next steps that need to be taken to towards realizing a ``defend forward" techniques in cyberspace.

    \begin{itemize}
        \item Further the fundamental research: While we have made considerable  advances in the fundamental research that forms the core of the data analysis engine identified in Section~\ref{sec:analysis}  \cite{kepner2019streaming,kepner2020750000000002020,kepner2019hypersparse,kepner2020multi-temporal}, substantial work remains prior to having a deployable system. Developing broader heuristics that can capture signatures of adversarial activity as well as leveraging AI/ML techniques to learn adversarial AI/ML infrastructures, and evolution of adversarial activity are a few of the areas where algorithm development is needed.
        \item Complete analysis of design tradeoffs, beginning with considerations identified in Section~\ref{sec:challenges}. Primary amongst these are determining the number of data sources required to have true observe-pursue-counter OFTH capability. While additional data points from sources can help with the confidence in the detected events, studies into identifying the point of optimal returns have yet to begin. Further, algorithmic costs associated with correlating multiple data fields (source/destination IP, time) from disparate data sources also has to be looked into more rigorously.
        \item Extend support to observatories: Much of the gray cyberspace data currently collected and maintained are due to the dedicated focus of a small underfunded community of actors \cite{kepner2021zero}. Since access to these data sets is vital, it is critical that support for these collection capabilities be continued, and expanded where possible.
        \item Developing appropriate policy framework: There are several public policy questions identified in \cite{kepner2021zero} that need to be addressed prior to, or alongside the technology development that is needed to achieve defend forward capabilities.
    \end{itemize}
\section{Conclusion}
    This document specifies the framework of a privacy-aware cyber defense tool named CyPhER, that takes in cyber data from multiple sources and conducts multiple types of statistical analysis on the data to detect adversarial activity. The framework uses an Open Design architecture, where decisions on specific design choices are postponed until proper mission CONOPS are available.
    
    The architectural framework, as described in Section~\ref{sec:arch}, allows a system that takes in anonymized source and destination IP addresses from collection points for inferential and predictive analysis of adversarial activity, and has the ability to recommend COAs if needed through prescriptive analysis. CyPhER strives to conduct most of its analysis using anonymized data, using only targeted enrichment in cases where it is essential for predictive or prescriptive analysis. While the modular and extensible architecture described in this document is flexible, several tradeoff scenarios need to be considered, as described in Section~\ref{sec:challenges}.
\section*{Acknowledgment}
The authors wish to acknowledge the following individuals for their contributions and support: Daniel Andersen, William Arcand, David Bestor, William Bergeron, Ayd{\i}n Bulu{\c{c}}, Bob Bond, Chansup Byun, K Claffy, Timothy Davis, Paula Donovan, Jeff Gottschalk, Micheal Houle, Matthew Hubbell, Anna Klein, Sarah McGuire, Lauren Milechin, Julie Mullen, Andrew Prout, Albert Reuther, Steve Rejto, Antonio Rosa, Siddharth Samsi, Charles Yee, and Marc Zissman

\bibliographystyle{IEEEtran}  
\bibliography{zot}

\begin{thebibliography}{10}
\providecommand{\url}[1]{#1}
\csname url@samestyle\endcsname
\providecommand{\newblock}{\relax}
\providecommand{\bibinfo}[2]{#2}
\providecommand{\BIBentrySTDinterwordspacing}{\spaceskip=0pt\relax}
\providecommand{\BIBentryALTinterwordstretchfactor}{4}
\providecommand{\BIBentryALTinterwordspacing}{\spaceskip=\fontdimen2\font plus
\BIBentryALTinterwordstretchfactor\fontdimen3\font minus
  \fontdimen4\font\relax}
\providecommand{\BIBforeignlanguage}[2]{{%
\expandafter\ifx\csname l@#1\endcsname\relax
\typeout{** WARNING: IEEEtran.bst: No hyphenation pattern has been}%
\typeout{** loaded for the language `#1'. Using the pattern for}%
\typeout{** the default language instead.}%
\else
\language=\csname l@#1\endcsname
\fi
#2}}
\providecommand{\BIBdecl}{\relax}
\BIBdecl

\bibitem{deeks2004defend}
A.~Deeks, ``Defend {Forward} and {Cyber} {Countermeasures},'' \emph{Ashley
  Deeks, Defend Forward and Cyber Countermeasures, Hoover Working Group on
  National Security, Technology, and Law, Aegis Series Paper}, no. 2004, 2020.

\bibitem{kosseff2019contours}
J.~Kosseff, ``The {Contours} of ‘{Defend} {Forward}’ {Under}
  {International} {Law},'' in \emph{2019 11th {International} {Conference} on
  {Cyber} {Conflict} ({CyCon})}, vol. 900.\hskip 1em plus 0.5em minus
  0.4em\relax IEEE, 2019, pp. 1--13.

\bibitem{montgomery2020cyberspace}
M.~Montgomery, B.~Jensen, E.~Borghard, J.~Costello, V.~Cornfeld, C.~Simpson,
  and B.~Valeriano, ``Cyberspace {S}olarium {C}ommission {R}eport,'' 2020.

\bibitem{gold1969digital}
B.~Gold and C.~M. Rader, ``Digital {P}rocessing of {S}ignals,'' \emph{Digital
  processing of signals}, 1969.

\bibitem{kay2013fundamentals}
S.~M. Kay, \emph{Fundamentals of {Statistical} {Signal} {Processing}:
  {Practical}{Algorithm}{Development}}.\hskip 1em plus 0.5em minus 0.4em\relax
  Pearson Education, 2013, vol.~3.

\bibitem{oppenheim1999discrete}
A.~V. Oppenheim, \emph{Discrete-time {Signal} {Processing}}.\hskip 1em plus
  0.5em minus 0.4em\relax Pearson Education India, 1999.

\bibitem{endsley1995toward}
M.~R. Endsley, ``Toward a {T}heory of {S}ituation {A}wareness in {D}ynamic
  {S}ystems,'' \emph{Human factors}, vol.~37, no.~1, pp. 32--64, 1995.

\bibitem{barford2010cyber}
P.~Barford, M.~Dacier, T.~G. Dietterich, M.~Fredrikson, J.~Giffin, S.~Jajodia,
  S.~Jha, J.~Li, P.~Liu, P.~Ning \emph{et~al.}, ``Cyber sa: Situational
  {A}wareness for {C}yber {D}efense,'' in \emph{Cyber situational
  awareness}.\hskip 1em plus 0.5em minus 0.4em\relax Springer, 2010, pp. 3--13.

\bibitem{chernyak2009brief}
V.~S. Chernyak and I.~Y. Immoreev, ``A brief history of radar,'' \emph{IEEE
  Aerospace and Electronic Systems Magazine}, vol.~24, no.~9, pp. B1--B32,
  2009.

\bibitem{kepner2016mathematical}
J.~Kepner, P.~Aaltonen, D.~Bader, A.~Bulu{\c{c}}, F.~Franchetti, J.~Gilbert,
  D.~Hutchison, M.~Kumar, A.~Lumsdaine, H.~Meyerhenke \emph{et~al.},
  ``Mathematical {F}oundations of the {GraphBLAS},'' in \emph{2016 IEEE High
  Performance Extreme Computing Conference (HPEC)}.\hskip 1em plus 0.5em minus
  0.4em\relax IEEE, 2016, pp. 1--9.

\bibitem{kepner2019new}
J.~Kepner, K.~Cho, and K.~Claffy, ``New {P}henomena in {L}arge-{S}cale
  {I}nternet {T}raffic,'' \emph{arXiv preprint cs.NI/1904.04396}, 2019.

\bibitem{kepner2021zero}
J.~Kepner, J.~Bernays, S.~Buckley, K.~Cho, C.~Conrad, L.~Daigle, K.~Erhardt,
  V.~Gadepally, B.~Greene, M.~Jones, R.~Knake, B.~Maggs, P.~Michaleas,
  C.~Meiners, A.~Morris, A.~Pentland, S.~Pisharody, S.~Powazek, A.~Prout,
  P.~Reiner, K.~Suzuki, K.~Takahashi, T.~Tauber, L.~Walker, and D.~Stetson,
  ``Zero {B}otnets: {A}n {O}bserve-{P}ursue-{C}ounter {A}pproach,''
  \emph{Belfer Center Reports}, vol.~6, 2021.

\bibitem{gortney2021dictionary}
W.~E. Gortney, ``{DOD} {D}ictionary of {M}ilitary and {A}ssociated {T}erms,''
  2012.

\bibitem{hutchins2011intelligence}
E.~M. Hutchins, M.~J. Cloppert, and R.~M. Amin, ``Intelligence-driven
  {C}omputer {N}etwork {D}efense {I}nformed by {A}nalysis of {A}dversary
  {C}ampaigns and {I}ntrusion {K}ill {C}hains,'' \emph{Leading Issues in
  Information Warfare \& Security Research}, vol.~1, no.~1, p.~80, 2011.

\bibitem{bessen2016knowledge}
J.~Bessen and A.~Nuvolari, ``Knowledge {S}haring among {I}nventors: {S}ome
  {H}istorical {P}erspectives,'' \emph{Revolutionizing Innovation: Users,
  Communities and Open Innovation. MIT Press: Cambridge, MA}, 2016.

\bibitem{howard2012open}
T.~J. Howard, S.~Achiche, A.~{\"O}zkil, T.~C. McAloone \emph{et~al.}, ``Open
  {D}esign and {C}rowdsourcing: {M}aturity, {M}ethodology and {B}usiness
  {M}odels,'' in \emph{DS 70: Proceedings of DESIGN 2012, the 12th
  International Design Conference, Dubrovnik, Croatia}, 2012, pp. 181--190.

\bibitem{pearce2012new}
J.~Pearce, S.~Albritton, G.~Grant, G.~Steed, and I.~Zelenika, ``A {N}ew {M}odel
  for {E}nabling {I}nnovation in {A}ppropriate {T}echnology for {S}ustainable
  {D}evelopment,'' \emph{Sustainability: Science, Practice and Policy}, vol.~8,
  no.~2, pp. 42--53, 2012.

\bibitem{sawant2013big}
N.~Sawant and H.~Shah, ``Big {D}ata {A}pplication {A}rchitecture,'' in
  \emph{Big data Application Architecture Q \& A}.\hskip 1em plus 0.5em minus
  0.4em\relax Springer, 2013, pp. 9--28.

\bibitem{kepner202075}
J.~Kepner, T.~Davis, C.~Byun, W.~Arcand, D.~Bestor, W.~Bergeron, V.~Gadepally,
  M.~Hubbell, M.~Houle, M.~Jones, A.~Klein, P.~Michaleas, L.~Milechin,
  J.~Mullen, A.~Prout, A.~Rosa, S.~Samsi, C.~Yee, and A.~Reuther,
  ``75,000,000,000 {S}treaming {I}nserts/{S}econd using {H}ierarchical
  {H}ypersparse graph{BLAS} {M}atrices,'' in \emph{2020 IEEE International
  Parallel and Distributed Processing Symposium Workshops (IPDPSW)}.\hskip 1em
  plus 0.5em minus 0.4em\relax IEEE, 2020, pp. 207--210.

\bibitem{graphblas}
\BIBentryALTinterwordspacing
``{GraphBLAS} {Repo}.'' [Online]. Available: \url{https://graphblas.github.io}
\BIBentrySTDinterwordspacing

\bibitem{command2018achieve}
U.~C. Command, ``Achieve and {Maintain} {Cyberspace} {Superiority}: {Command}
  {Vision} for {US} {Cyber} {Command},'' \emph{https://bit.ly/3i0HWEI}, vol.
  202018, 2018.

\bibitem{smeets2020us}
M.~Smeets, ``{US} {Cyber} {Strategy} of {Persistent} {Engagement} \& {Defend}
  {Forward}: {Implications} for the {Alliance} and {Intelligence}
  {Collection},'' \emph{Intelligence and National Security}, vol.~35, no.~3,
  pp. 444--453, 2020, publisher: Taylor \& Francis.

\bibitem{kepner2019streaming}
J.~Kepner, V.~Gadepally, L.~Milechin, S.~Samsi, W.~Arcand, D.~Bestor,
  W.~Bergeron, C.~Byun, M.~Hubbell, M.~Houle, M.~Jones, A.~Klein, P.~Michaleas,
  J.~Mullen, A.~Prout, A.~Rosa, C.~Yee, and A.~Reuther, ``Streaming 1.9
  {Billion} {Hypersparse} {Network} {Updates} per {Second} with {D4M},'' in
  \emph{2019 {IEEE} {High} {Performance} {Extreme} {Computing} {Conference}
  ({HPEC})}, 2019, pp. 1--6.

\bibitem{kepner2020750000000002020}
\BIBentryALTinterwordspacing
J.~Kepner, T.~Davis, C.~Byun, W.~Arcand, D.~Bestor, W.~Bergeron, V.~Gadepally,
  M.~Hubbell, M.~Houle, M.~Jones, A.~Klein, P.~Michaleas, L.~Milechin,
  J.~Mullen, A.~Prout, A.~Rosa, S.~Samsi, C.~Yee, and A.~Reuther,
  \emph{75,000,000,000 {Streaming} {Inserts}/{Second} {Using} {Hierarchical}
  {Hypersparse} {GraphBLAS} {Matrices}}, 2020, \_eprint: 2001.06935. [Online].
  Available: \url{https://arxiv.org/abs/2001.06935}
\BIBentrySTDinterwordspacing

\bibitem{kepner2019hypersparse}
\BIBentryALTinterwordspacing
J.~Kepner, K.~Cho, K.~Claffy, V.~Gadepally, P.~Michaleas, and L.~Milechin,
  ``Hypersparse {Neural} {Network} {Analysis} of {Large}-{Scale} {Internet}
  {Traffic},'' \emph{2019 IEEE High Performance Extreme Computing Conference
  (HPEC)}, Sep. 2019, iSBN: 9781728150208 Publisher: IEEE. [Online]. Available:
  \url{http://dx.doi.org/10.1109/HPEC.2019.8916263}
\BIBentrySTDinterwordspacing

\bibitem{kepner2020multi-temporal}
J.~Kepner, C.~Meiners, C.~Byun, S.~McGuire, T.~Davis, W.~Arcand, J.~Bernays,
  D.~Bestor, W.~Bergeron, V.~Gadepally, and {others}, ``Multi-{Temporal}
  {Analysis} and {Scaling} {Relations} of 100,000,000,000 {Network}
  {Packets},'' \emph{arXiv preprint arXiv:2008.00307}, 2020.

\end{thebibliography}
\end{document}